\documentclass[aps,prb,twocolumn,showpacs]{revtex4-2}

\usepackage{amsfonts}
\usepackage{braket}
\usepackage{mathrsfs}
\usepackage{xcolor}
\definecolor{myOrange}{rgb}{1,0.5,0}

\pdfoutput=1
\usepackage{amsmath}
\usepackage{bm}
\usepackage{amssymb}
\usepackage{graphicx}
\usepackage{color}
\usepackage{simplewick}
\usepackage[version=3]{mhchem}
\usepackage{wasysym}
\usepackage{physics}
\usepackage{siunitx}
\usepackage{hyperref}
\usepackage[thinc]{esdiff}

\DeclareMathOperator{\sign}{sign}

\newcommand{\magvec}{\vb{M}}
\newcommand{\magvecunit}{\hat{\vb{M}}}

\renewcommand{\vec}[1]{{\boldsymbol #1}}

\definecolor{paperblue}{rgb}{0.0078,.133,.749}

\begin{document}

\title{Fractional Topological Charges in Two-Dimensional Magnets}

\newcommand{\colt}{Institute for Theoretical Physics,
University of Cologne, 50937 Cologne, Germany}

\author{Nina del Ser}
\affiliation{\colt}
\author{Imane El Achchi}
\affiliation{\colt}
\author{Achim Rosch}
\affiliation{\colt}

\date{\today}



\begin{abstract}
Magnetic skyrmions and antiskyrmions are characterised by an integer topological charge $\mathcal Q =\mp 1$,  describing the winding of the magnetic orientation. Half-integer winding numbers, $\mathcal Q=\pm \frac{1}{2}$, can be obtained for magnetic vortices (merons). Here, we discuss the physics of magnets with fractional topological charge which is neither integer nor half-integer. We argue that in ferromagnetic films with cubic anisotropy, textures with $\mathcal Q=\pm\frac{1}{6} $ or $\pm\frac{1}{8}$ arise naturally when three or more magnetic domains meet. We also show that a single magnetic skyrmion with $\mathcal Q =-1$ can explode into four fractional defects, each carrying charge $\mathcal Q=-\frac{1}{4}$. Additionally, we investigate a  point defect with a non-quantised fractional charge ($\mathcal Q\neq \frac{n}{m}, n,m\in\mathbb{Z}$) which can move parallel to a magnetic domain wall. Only  defects with fractional charge lead to an Aharonov-Bohm effect for magnons. 
We investigate the resulting forces on a fractional defect due to magnon currents. 

 \end{abstract}


\maketitle



The topology of magnetic textures is one of their most important defining features \cite{kosevichReview1990,nagaosa2013topological,BraunReview2012,reviewEverschor,GoebelBeyondSkyrmionsReview2021}.
Magnetic domain walls, vortices and merons, skyrmions, hopfions, and Bloch points are some intensively studied examples of
topological excitations in magnets \cite{GoebelBeyondSkyrmionsReview2021,petti2022review}.

Magnetic skyrmions \cite{muhlbauer2009skyrmion,Yu2010RealspaceOO,heinze2011,nagaosa2013topological,jonietz2010spin,Schulz2012,BraunReview2012,reviewEverschor}, for example, are characterised by an 
integer winding number: the orientation $\magvecunit(\vb{r})=\magvec(\vb{r})/|\magvec(\vb{r})|$ of the magnetisation $\magvec(\vb{r})$ winds once around the unit sphere.  Mathematically, the winding number $\mathcal Q$ of smooth two-dimensional magnetic textures is calculated from the topological density $\rho_{\text{top}}$, using
\begin{align}
\mathcal Q  =\frac{1}{4 \pi} \int \dd^2{r}  \rho_{\text{top}}(\vb{r}), \quad \rho_{\text{top}}(\vb{r})= \magvecunit\cdot (\nabla_x \magvecunit \times \nabla_y \magvecunit), \label{winding}
\end{align}
with $\mathcal Q=-1$ for a skyrmion embedded in a spin-up ferromagnet.
Here, $\int \dd^2{r}\rho_{\text{top}}(\vb{r})$ measures the solid angle covered by the  vectors $\magvecunit(\vb{r})$.
Any system which has periodic boundary conditions or is smoothly embedded into a ferromagnetic background will have an integer winding number, $\mathcal Q \in \mathbb Z$, as the total solid angle is a multiple of $4 \pi$. 

Non-integer winding numbers of point-like defects can be obtained by considering magnetic textures with non-trivial (non-ferromagnetic) orientations far away from the origin. A prime example are so-called merons \cite{kosevichReview1990,EzawaMeron2011,Yu2018,GoebelBeyondSkyrmionsReview2021}.
These are specific types magnetic vortices, realised, for example, in magnets with an easy-plane anisotropy. Far from the centre of the meron, the magnetisation winds around the equator of the unit sphere (thus the structure is a vortex). Close to the centre, the magnetisation points either up or down. 
 Therefore, $\magvecunit(\vb{r})$ covers half of the unit sphere, giving rise to a topological charge of either $\mathcal Q=+\frac{1}{2}$ or $\mathcal Q=-\frac{1}{2}$, respectively.
 Merons play an important role in the stability of magnetic skyrmions. 
When the stabilising magnetic field is lowered, a skyrmion embedded in a ferromagnetic background becomes unstable and may decay into a pair of merons (often called a bimeron), each of which carries $\mathcal Q=-\frac{1}{2}$, or exactly half of the skyrmion's original integer winding number $\mathcal Q=-1$ \cite{EzawaMeron2011,SchuetteScatt2014,ohara2022reversible,yu2024spontaneous}. 

Recently, another class of defects with fractional charge has also been considered: fractional charges which occur close to the boundary of the sample \cite{Loss2020,Loss2021,jena2022}. In this case, the absence of quantisation arises from the restricted area of integration in Eq.~\eqref{winding}. Another possibility is to form a lattice of fractional charges, where, however, the fractional charges have to add up to an integer when the total unit cell is considered
 \cite{gao2020,boemerich2020}.
In the quantum Hall phase of twisted bilayer graphene \cite{boemerich2020}, we found, for example, that gate voltages stabilise a so-called double tetarton lattice, consisting of topological defects covering one quarter of the unit sphere twice, but the charge per magnetic unit cell remains quantised to an integer. In Ref.~\cite{hayami2022}, a magnetic texture in a layered system was considered with either an integer or zero skyrmion winding number per layer. Averaging this over layers, one obtains a fractional number. We will not consider either of these cases in the following, instead focusing solely on isolated fractional charges which form in the bulk of magnetic films.

The presence of a topological winding number has profound effects  on the stability of magnetic textures, the possibilities to manipulate them and on their interaction with electrons and magnons. Skyrmions, for example, can change their winding number only via singular spin configurations \cite{Heil2019}. Perhaps more importantly, magnons and electrons pick up Berry phases when following a magnetic texture. The physical effect of the Berry phase is best described as an emergent electromagnetic field \cite{Volovik1987,Zhang2011,Schulz2012}, given by
\begin{align}
B_{\text{eff}}(\vb{r})=  \frac{\hbar}{e} \rho_{\text{top}}(\vb{r}). \label{Beff}
\end{align} 
Here, we use a convention where  majority or minority electrons carry an emergent charge $q=\mp \frac{1}{2}e$, respectively, while $q=e$ for magnons, which can be obtained by flipping a majority to a minority electron. The electron charge $e$ in these formulas has been introduced such that $B_{\text{eff}}$ has units of Tesla, but it has no physical meaning as it drops out of the product $q B_{\text{eff}} $. In units of the flux quantum $\Phi_0=\frac{2 \pi \hbar}{q}$, the total flux is  given by $\mp\mathcal Q$ for majority(minority) electrons and $2 \mathcal Q$ for magnons. Incoming magnons or electrons will generally be scattered by this emergent magnetic flux. However, a fundamental distinction between integer and fractional $\Phi$ occurs if the incoming particles have low energy and momentum $k$. In this case, Aharonov and Bohm \cite{AharanovBohm1959} showed that the scattering cross-sections drop to zero for integer $\Phi$, because such a flux can be removed by a gauge transformation. For fractional $\mathcal Q$, however, the cross section grows {\em singularly} with $1/k$.


In the following, we will first point out that fractional topological textures occur frequently in standard ferromagnetic systems and give several examples. Then, we will show how a skyrmion may be triggered to explode into smaller fractionally charged topological defects. Finally, we will explore the scattering of magnons from fractional textures and the Aharonov-Bohm effect for magnons. 


\begin{figure}
\includegraphics{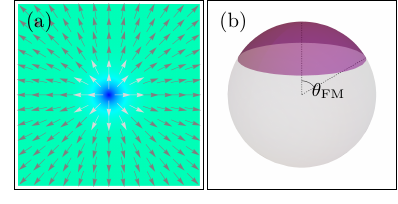}%
	\caption{Fractional vortex, modelled by Eq.~\eqref{eq:vortexEnergy} using $b/(2\kappa_u)=0.54$. Panel (a) shows the magnetisation in the 2D plane. The  colour encodes the $\hat{M}_z$ component (colour scale shown to the right of Fig.~\ref{fig:cubicFractionalCharges}), while arrows indicate the in-plane magnetisation direction $(\hat{M}_x,\hat{M}_y)^T$. The purple area in panel (b) shows the  area covered by $\magvecunit$ in spin space. The topological charge $\mathcal Q$ is the ratio of the purple area to the full surface area of the sphere $4 \pi$, in this case $\mathcal{Q}=0.23$.
	}\label{fig:vortexTopoCharge}%
\end{figure}

{\em Textures with fractional topological charge ---} Arguably the simplest way to generate a fractional topological charge $\mathcal Q$ is to consider a magnetic vortex in a two-dimensional easy-plane magnet in the presence of an out-of-plane magnetic field. The energy density of such a system is given by
\begin{equation}\label{eq:vortexEnergy}
\mathcal{F}=\frac{A}{2} (\nabla \magvecunit)^2+\kappa_{u} \hat{M}_z^2-b\hat{M}_z,
\end{equation}
where $A$ is the spin stiffness of the ferromagnet, $\kappa_u>0$ is the anisotropy energy density and $b=B |\magvec|$  parametrises the strength of the external magnetic field. Note that as we are in 2D, $\nabla=(\nabla_x,\nabla_y,0)^T$ throughout the paper.
Due to the interplay of anisotropy and magnetic field, the magnetisation in the ferromagnetic phase is tilted, and can lie anywhere on the cone defined by the angle 
$\theta_{\text{FM}}=\arccos[b/(2 \kappa_{u})]$, relative to the $z$-axis. In this case, magnetic vortices and antivortices have topological charge
\begin{align}
\mathcal Q=\pm \frac{1}{4 \pi} \int_0^{2 \pi}\dd{\phi} \int_0^{\theta_{\text{FM}}}\dd{\theta}\sin\theta =\pm \left(\frac{1}{2} - \frac{|b|}{4 \kappa_u}\right).
\end{align}
For $b=0$, one obtains merons with $\mathcal Q=\pm 1/2$, whereas for finite $|b|<2 \kappa_u$, the charge takes smaller values. We will argue below that 
the properties of such a state change qualitatively when $|\mathcal Q|$ is not a multiple of $1/2$. Recently, fractional vortices have also been investigated as so-called ``screw dislocations'' in helical magnets, see \cite{azhar2022} for some of their more exotic manifestations.

\begin{figure}
\includegraphics[width=0.38\textwidth]{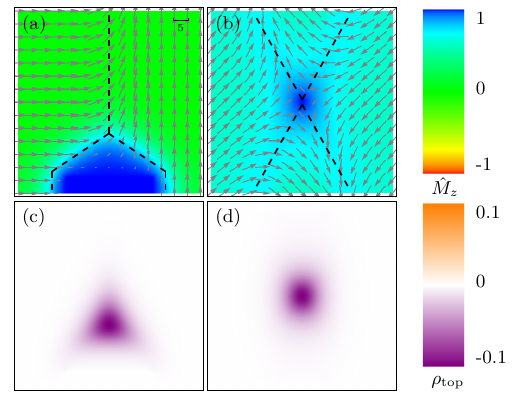}%
	\caption{
Fractional charges in magnets with cubic anisotropy. Panels (a) and (b) show $\magvecunit(\vb{r})$ in a 2D system where three or four domains meet, with $A=1$, $\kappa_c=\pm 0.05$, and lattice spacing $\Delta=0.1$, respectively (data obtained from micromagnetic simulations). Dashed lines show the 1D domain walls between each pair of domains. Panels (c) and (d) show the spatial distribution of the topological charge density $\rho_{\text{top}}(\vb{r})$.
	}\label{fig:cubicFractionalCharges}%
\end{figure} 


\textit{Cubic magnets ---}
For $b=0$, fractional charges arise naturally in two-dimensional ferromagnets  with cubic anisotropy 
described  by the energy density
\begin{eqnarray}
\mathcal{F}=\frac{A}{2} (\nabla \magvecunit)^2-\kappa_c (\hat{M}_x^4+\hat{M}_y^4+\hat{M}_z^4). \label{eq:cubicMagnetHamiltonian}
\end{eqnarray}
For $\kappa_c>0$, there are six domains (easy axes) with orientation $\magvecunit_{\text{FM}} \in \{(\pm 1,0,0)^T,(0,\pm 1,0)^T,(0,0,\pm 1)^T\}$.
At a $90^\circ$ domain wall, e.g.  from $(1,0,0)^T$ to $(0,1,0)^T$, the magnetisation interpolates preferentially via the medium axis $(1,1,0)^T/\sqrt{2}$, keeping $\hat{M}_z=0$ throughout. This is guaranteed by the product of time-reversal symmetry and a $180^\circ$ rotation symmetry about the $z$-axis. Thus, it can be parametrised by $\magvecunit(x)=(\cos(\phi(x)),\sin(\phi(x)),0)^T$, with $\phi(x)=\arctan(e^{\sqrt{\frac{4 \kappa_c}{A}} (x-x_0)})$, the well-known soliton of the sine-Gordon equation \cite{gurevich2008,wheeler2015}. A 180$^\circ$ domain wall, e.g. between $(1,0,0)^T$ and $(-1,0,0)^T$, is not stable in our model and decays into two 90$^\circ$ domain walls. In two-dimensional systems, three domains generically meet at a point. The topological charge at the meeting point of the three 90$^\circ$ domains can easily be obtained by tracking the magnetisation far away from the meeting point. As an example, consider a 2D system consisting of three magnetic domains $\magvecunit_{\text{FM}1}=(0,0,1)^T$, $\magvecunit_{\text{FM}2}=(0,1,0)^T$ and $\magvecunit_{\text{FM}3}=(1,0,0)^T$. 
The three domains smoothly transform into each other as we move anticlockwise around their common meeting point. Far away from the defect, the magnetisation interpolates exactly along the great arc connecting the domain orientations, as discussed above.
This implies that the magnetisation covers exactly $1/8$ of the unit sphere and thus 
\begin{align}
\mathcal Q=\pm \frac{1}{8} .
\end{align}
\begin{figure}
\includegraphics[width=0.47\textwidth]{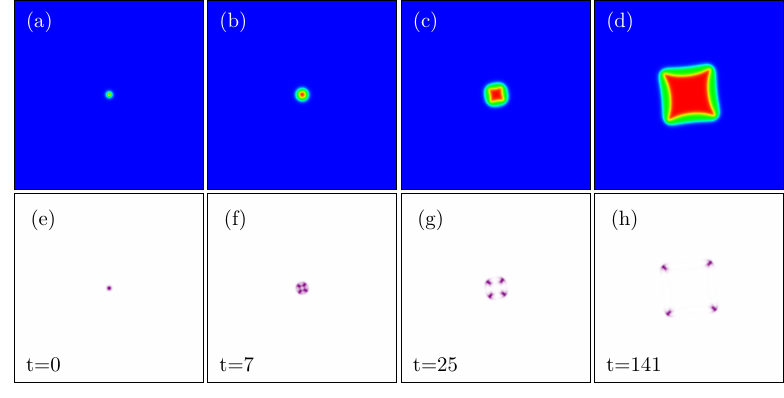}%
	\caption{
A magnetic skyrmion exploding into four identical fractional charges with topological charge $\mathcal Q=-\frac{1}{4}$ each, for a system with positive anisotropy prefactor, $\kappa_c>0$ (model: Eq.~\eqref{eq:skyrmionAnistoropiyFreeEnergy} with parameters $A=1$, $D=1$, $b=0.8$, $\kappa_c=3.2$, $\Delta=0.1$, the results depends only on the ratios $ A \kappa_c/D^2$ and $A b/D^2$). Panels (a)--(d) show the magnetisation in the $z$-direction. Panels (e)--(h) show the distribution of the topological charge. Colour scales for $\hat{M}_z$ and $\rho_{\text{top}}$ are as in Fig.~\ref{fig:cubicFractionalCharges}. Data generated by quenching from $b_{\text{ext}}=0.8$ to $b_{\text{ext}}=0$ at $t=0$, keeping $\kappa_c=3.2$ constant throughout the simulation. 
	}\label{fig:decayingSkyrmion}
\end{figure}
Note that this value is topologically quantised: it is robust against all local deformations. To change it, one has to either modify the spin orientation within the domain (at an energy cost proportional to $L^2$, where $L$ is the system size), or one has to change how the system interpolates from one domain to the next in the domain walls (at an energy cost proportional to $L$).
The sign of $\mathcal Q$ is determined by the ordering of the domains, $\sign \mathcal Q=\sign[\magvecunit_{\text{FM1}}\cdot ( \magvecunit_{\text{FM2}} \times \magvecunit_{\text{FM3}})]$.
In Fig.~\ref{fig:cubicFractionalCharges}(a), we show a micromagnetic simulation performed using mumax3 \cite{vansteenkiste2014} of such a texture. We use open boundary conditions and fix the orientation of a few spins along the edge of the simulated region in the relevant direction to encourage the formation of each domain. The plot of the topological charge density, Fig.~\ref{fig:cubicFractionalCharges}(c), shows that the charge is well-localised around the defect centre on a length scale set by the typical width of domain walls, $\sqrt{A/\kappa_c}$. Similarly, for $\kappa_c<0$, one naturally obtains domains with $\mathcal Q=\pm \frac{1}{6}$, see App.~\ref{app:CubicMagnetsNegativeKappaC}. In this case, four domains meet, see Fig.~\ref{fig:cubicFractionalCharges}(b). The discussion above ignores dipolar interactions, which break the cubic symmetry of a magnetic film.
In this case, the ``quantisation'' to simple fractions like $1/8$ or $1/6$ is only approximate.
An alternative way to classify such  defects stabilised by anisotropies has recently been suggested by Rybakov and Eriksson \cite{Rybakov2022}. They view them as (non-abelian) vortices whose energetics (and  topological classification by the first homotopy group) is controlled by avoiding the relevant magnetic hard axes.

\textit{Exploding skyrmion ---} Another way to generate quantised fractionally charged defects is via skyrmion explosion. To show this, we consider the model
\begin{align}
\mathcal{F}=&\frac{A}{2} (\nabla \magvecunit)^2+D(\hat{M}_z(\nabla\cdot \magvecunit )-(\magvecunit\cdot\nabla)\hat{M}_z)  \nonumber \\
&-\kappa_c (\hat{M}_x^4+\hat{M}_y^4+\hat{M}_z^4) -b(t) \hat{M}_z \label{eq:skyrmionAnistoropiyFreeEnergy}
\end{align}
in a regime where a single magnetic skyrmion is  stabilised by an external magnetic field $b$. Here, $D$ parametrises the strength of Dzyaloshinskii-Moriya interactions (DMI). We choose $\kappa_c=3.2$, such that anisotropy fields are sufficiently large to suppress helical phases at $b=0$. Suddenly switching off the external magnetic field triggers the skyrmion with $\mathcal Q=-1$ to ``explode'' into four defects with charge $\mathcal Q=-\frac{1}{4}$. The defects form at the corners of a square, where two domain walls meet and end, see Fig.~\ref{fig:decayingSkyrmion}. In the chosen example, the domain walls have a negative energy per length and provide the energy gain to drive the fractional defects apart. We also tried the same experiment with $\kappa_c<0$. In this case, however, the ferromagnetic $(0,0,1)$ state far from the skyrmion is unstable as the anisotropy term favours the $(\pm 1,\pm 1,\pm 1)/\sqrt{3}$ directions. Numerically, we obtain a proliferation of more and more defects with charge $\mathcal{Q}=\frac{1}{6}$, although there are always six more negative ones than positive ones to conserve the initial $\mathcal Q$, see App.~\ref{app:skyrmionExplosionNegativeAnistropy}.

\begin{figure}
\includegraphics[width=.45\textwidth]{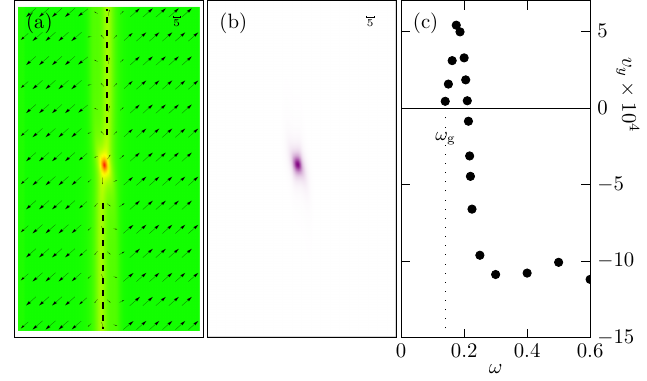}%
	\caption{Panels (a) and (b) show the magnetisation and topological charge distribution of the fractional defect formed between two symmetry-broken chiral domain walls, respectively (model: Eq.~\eqref{eq:modelEasyPlaneChiralMagnet} with $A=1$, $A_\text{ani}=10$, $D=0.09$, $\kappa_p=0.05$, $\kappa_u=0.1$, $b=-0.006$, $\Delta=1.25$, distance between left edge and defect $L_x=60$, see Fig.~\ref{fig:cubicFractionalCharges} for colour scales). Total (non-quantised) charge $\mathcal{Q}=-0.352$. A magnon current is induced by oscillating the spins on the left edge with frequency $\omega$. This leads to a net motion of the defect parallel to the domain wall. The resulting velocity $v_y$ is plotted as a function of $\omega$ in panel (c) for induced magnetisation strength $\delta M/M_0=0.0125$ and damping $\alpha=0.03$. 
 The sign change of $v_y$ arises as the Lorentz force and the Aharonov-Bohm effect lead to opposite signs of the transverse cross section $\sigma_\perp$.
	}\label{fig:easyplaneDMI}%
\end{figure}

\textit{Mobile fractional defects within symmetry-broken domain walls ---}
A different class of defects can be obtained by investigating topological textures occurring as a ``domain wall within a domain wall''. Here, we consider a 
2D system with easy-plane anisotropy terms, an external magnetic field and DMI,
\begin{align}
\mathcal{F}&=\frac{A}{2} (\nabla \magvecunit)^2+\frac{A_{\text{ani}}}{2}(\nabla_y \magvecunit)^2 +\kappa_p (\hat{M}_x^4+\hat{M}_y^4) \nonumber \\
&+ \kappa_{u}\hat{M}_z^2 -b \hat{M}_z+D(\hat{M}_z(\nabla\cdot \magvecunit )-(\magvecunit\cdot\nabla)\hat{M}_z), \label{eq:modelEasyPlaneChiralMagnet}
\end{align}
where, to help stabilise the system once magnon currents are added, we also added some anisotropy to the Heisenberg coupling by introducing $A_{\text{ani}}$ (see App.~\ref{app:NumericsDefect} for further details).
In the case $\kappa_p,\kappa_u>0$, Eq.~\eqref{eq:modelEasyPlaneChiralMagnet} has four degenerate domains in the ferromagnetic phase, parametrised by $\phi_{\text{FM}}=\frac{\pi}{4}+n\frac{\pi}{2}, n\in \{0,1,2,3\}$ and
\begin{equation}
\begin{aligned}
\theta_{\text{FM}}&=\begin{cases}
\theta_{\text{FM}}(b,\kappa_p,\kappa_u), & 0\leq |b|\leq 2\kappa_u, \\
\frac{\pi}{2}(1-\sign(b)), & |b|>2\kappa_u,
\end{cases}
\end{aligned}
\end{equation}
where $\theta_{\text{FM}}(b,\kappa_p,\kappa_u)$ is defined in App.~\ref{app:DomainWalls}. Denoting the magnetisation perpendicular to the magnetic field by $n_\perp$, in Fig.~\ref{fig:easyplaneDMI}(a) we consider a domain wall between two phases with $n_\perp \sim (-1,-1)^T$ and $n_\perp \sim (1,1)^T$ along the left and right edges of the system, respectively. Inside the domain wall, the magnetisation can interpolate via either the $(1,-1)^T$  or the $(-1,1)^T$ direction. The system chooses one of these options spontaneously, defining two symmetry-related types of domain walls.  Fig.~\ref{fig:easyplaneDMI}(a) shows a case where one solution is realised in the top and the other in the bottom part of the magnetic texture. Thus, one obtains a defect with a fractional charge which can be viewed as a zero-dimensional domain wall within the domain wall separating the two domains.
$\mathcal Q$ is not quantised and can be controlled by the magnetic field.
This fractional defect has a property which makes it different from all the others discussed before: it can move at zero energy cost parallel to the domain wall using only {\em local} changes of the magnetisation. In contrast, for all the cases previously considered, a translational motion of fractional defects required changes of the magnetisation far away from the defect.

{\em Dipolar Interactions -- }
In our discussion, we have ignored the role of long-ranged dipolar interactions. This is justified as for thinner and thinner films, the long-ranged part of the dipolar interactions becomes less important, see, e.g., Ref. \cite{gioia1997micromagnetics}, and the only surviving effect is a change of the value of the  anisotropy term $\kappa_u$. In our discussion of cubic magnets, Eq.~\eqref{eq:cubicMagnetHamiltonian} and  Eq.~\eqref{eq:skyrmionAnistoropiyFreeEnergy}, we have implicitly assumed that the renormalised value of $\kappa_u$ is sufficiently small to be ignored.
As noted above, the quantisation of the topological charge to fractional values like $1/8$ or $1/6$ can be affected by a finite $\kappa_u$.

{\em Aharonov-Bohm scattering ---} What distinguishes defects with a fractional charge from skyrmions (and merons) with an integer (half-integer) winding number? A key difference is the Aharonov-Bohm effect. The standard Aharonov-Bohm effect describes how a charged particle picks up a quantum-mechanical phase when encircling 
a magnetic flux, despite never coming in contact with the magnetic field. 
As discussed in the introductory section,  a topological defect with charge $\mathcal Q$ acts like an effective magnetic field with flux $2 \mathcal Q \Phi_0$, where $\Phi_0$ is the flux quantum. If $2 \mathcal Q$ is integer, there is no Aharonov-Bohm effect: an integer flux can be gauged away. In contrast, for fractional $\mathcal Q$, one obtains, for $k\to 0$, a highly singular differential cross section, as first calculated by Aharonov and Bohm \cite{AharanovBohm1959}.

One way to observe this effect is to track the force on the defect in the presence of a magnon current $j_m$. The force perpendicular to $j_m$ is computed (in the limit of weak damping) from the change of magnon momentum due to scattering \cite{lekner2005},
\begin{align}\label{eq:forceTransverseMagnonCurrent}
 F_\perp =\hbar k \, j_m\, \sigma_\perp, \qquad \sigma_\perp={ \int}\mathrm{d}\chi(-\sin\chi)\frac{\mathrm{d}\sigma}{\mathrm{d}\chi},
\end{align}
where $\frac{\mathrm{d}\sigma}{\mathrm{d}\chi}$ is the differential cross section and $\hbar k$ is the magnon momentum. If the wavelength of magnons is short compared to the size $R_d$ of the defect, $k R_d \gg 1 $, the differential cross section can be calculated from the classical Lorentz force $q (\vb v \times \vb B)$, while for $k R_d \ll 1$ we use the Aharonov-Bohm result \cite{AharanovBohm1959}, see App.~\ref{sec:scattering}. Therefore, we approximate
\begin{align}
\sigma_{\perp}& \approx \left\{ \begin{array}{cc}
      \frac{\sin( 4 \pi \mathcal Q)}{ k}  T_k  &  \quad {\text{for}} \quad k R_d \ll 1,\\[2mm]
    \frac{4 \pi \mathcal Q}{ k} &  \quad {\text{for}} \quad  k R_d \gg 1, 
\end{array} \right.  \label{eq:sigmaLimits}
\end{align}
where the transmission rate $T_k$, $0\le T_k \le 1$, is a correction factor which we add to take into account that only those magnons which are transmitted through the domain walls contribute to Aharonov-Bohm scattering, see App.~\ref{sec:scattering} for details. While the Aharonov-Bohm effect is periodic in the flux, the Lorentz force is simply linear in $\mathcal Q$.
The smooth cross-over between the two limits has been discussed in the PhD thesis of Vivek Lohani \cite{lohani2022}. Importantly, the sign of $\sigma_\perp$ changes as a function of $k$ for $1/2 <2 \mathcal |Q| < 1$. The Aharonov-Bohm effect and Lorentz force have opposite signs in this range of $|\mathcal{Q}|$, and thus low- and high-energy magnons scatter in opposite directions. This effect will occur even if $T_k$ becomes small for small $k$.
Therefore, we can use a sign change in the force as a smoking gun signature of  Aharonov-Bohm scattering, which is only activated for fractionally charged defects.
Fig.~\ref{fig:easyplaneDMI}(c) shows the result of numerical simulations for the ``domain wall in a domain wall'' defect shown in Fig.~\ref{fig:easyplaneDMI}(a), where a magnon current is generated by inducing an oscillating magnetisation $\delta \vb{M}(t)$ along the left edge of the simulated system. The force arising from the resulting magnon current induces a motion of the defect perpendicular to the magnon current (and therefore parallel to the domain wall) with velocity $v_\perp \propto F_\perp$. The velocity shows the expected sign change when one crosses over from the Aharonov-Bohm regime  (driving frequency $\omega$ close to the gap of the ferromagnet) to the Lorentz-force regime (larger $\omega$).


{\em Conclusions ---} In our paper, we have argued that magnetic textures with fractional topological charge are not rare: they naturally occur, for example, in magnetic vortices in a transverse magnetic field or when three ferromagnetic domain walls meet in systems with cubic anisotropy. Additionally, if a magnetic skyrmion splits into four pieces upon lowering the magnetic field, each piece obtains $1/4 $ of the initial topological charge. Finally, we discussed a mobile fractionally charged defect arising from a domain wall within a domain wall. In some of the cases above, the fractional topological charge took rational values like $1/6$ or $1/8$, which was guaranteed by additional symmetries.

Two related features distinguish fractional topological textures from their integer counterparts. First, the magnetisation far away from a fractional defect is necessarily non-collinear. For example, several domain walls meet at the location of the defect. Second, magnons pick up a Berry phase $4 \pi \mathcal Q $ when moving around the defect, giving rise to an Aharonov-Bohm effect when $\mathcal Q$ is neither integer nor half-integer. This phase is picked up when the magnon traverses the domain walls. The interference of magnons can then lead to an effective {\em Aharonov-Bohm force} on the defect.
Here, it is most interesting to consider defects with $\frac{1}{4}<\mathcal |Q| < \frac{1}{2}$, as in this case Lorentz and Aharonov-Bohm forces have opposite signs.

For the future, it will be interesting to investigate the coupled dynamics of magnons and fractionally charged textures. How can  one obtain bound states or lattices of fractionally charged defects and how do they interact with magnon currents? We expect that the fractional emergent flux and the resulting Aharonov-Bohm effect will play an important role in such settings. An exciting option is also to use modern space-resolved imaging techniques \cite{Girardi2024} to directly map in real space how fractional topological textures can deflect spin currents.

{\em Acknowledgements ---} We thank Maria Azhar, Volodymyr Kravchuk, and Markus Garst for useful discussions. Special thanks to Vivek Lohani for his insights and results on Aharonov-Bohm scattering.
We acknowledge the financial support of
CRC 1238 (project number 277146847, subproject C04).
We also thank the Regional Computing Center of the University of Cologne (RRZK) for providing computing time on the DFG-funded (funding number: INST 216/512/1FUGG) High Performance Computing (HPC) system CHEOPS as well as technical support. Finally, we thank Andreas Sindermann and Octavio del Ser for valuable technical support.



\bibliography{Library}

\appendix

\section{Domains with charge $\pm\frac{1}{6}$}\label{app:CubicMagnetsNegativeKappaC}
\begin{figure}
\includegraphics{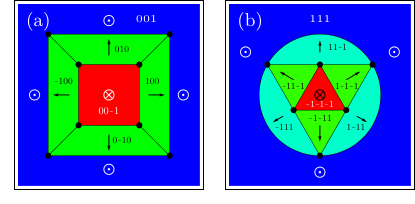}%
	\caption{
Sketch of domain wall configurations in a cubic magnet. Both configurations have a quantised total winding number, $\mathcal Q=-1$ (characteristic for a magnetic skyrmion).  Panel (a): for $\kappa_c>0$, there are six domains. Stable domain walls exist when neighbouring domains are perpendicular to each other. This results in eight topological defects (black dots), where three domain walls meet. 
As they carry the same topological charge by symmetry, one obtains $\mathcal Q=-\frac{1}{8}$ per defect. Panel (b): a similar scheme can be used for  $\kappa_c<0$, where eight domains exist. In this case, four domain walls meet at a defect, see Fig.~\ref{fig:cubicFractionalCharges}(b). There are six point defects, each carrying $\mathcal Q=-\frac{1}{6}$. Colour denotes the angle between the outer and local domain orientations, while the arrows show a projection of the magnetisation in the plane perpendicular to the magnetisation of the outer domain.
	}\label{fig:explodedSkyrmionTriangleSquare}%
\end{figure}

For a ferromagnet with negative cubic anisotropy, $\kappa_c<0 $ in Eq.~\eqref{eq:cubicMagnetHamiltonian}, eight magnetic domains with orientation  $\magvecunit_{\text{FM}}=(\pm 1,\pm 1,\pm 1)^T/\sqrt{3}$ minimise the energy. The spins in these domains point towards the eight corners of the cube, see Fig.~\ref{fig:cubicFractionalCharges}(b). As we have checked numerically, only the domains connecting nearest-neighbour corners on the cube are stable. Therefore, the only stable point-like configuration occurs when four domains belonging to the same face of the cube meet, forming a  vortex-like (or anti-vortex like) configuration, see Fig.~\ref{fig:cubicFractionalCharges}(b). 
Due to the cubic symmetry, these defects have the same topological charge (up to a sign). Furthermore, one can combine six defects with the same winding number to a skyrmion with winding number $\mathcal Q=-1$, see Fig.~\ref{fig:explodedSkyrmionTriangleSquare}. Thus, each of these defects has to have topological charge 
\begin{align}
\mathcal Q=\pm \frac{1}{6}.
\end{align}

\section{Skyrmion Explosion for negative $\kappa_c$ }\label{app:skyrmionExplosionNegativeAnistropy}

\begin{figure}
\includegraphics[width=0.47\textwidth]{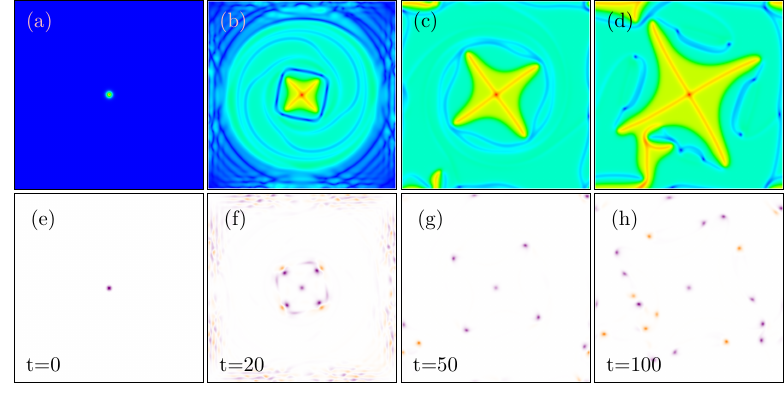}%
	\caption{Magnetic skyrmion explosion for the model in Eq.~(7) with negative $\kappa_c=-3.2$ (all other parameters same as in Fig.~3). Panels (a)-(d) show the $M_z$ component of the magnetisation, while panels (e)-(h) show $\rho_{\text{top}}(\vb{r})$ (colour scheme as in Fig.~\ref{fig:cubicFractionalCharges}). The background magnetisation $(0,0,1)^T$ is initially very unstable, see panel (b), as the background spins have the choice to decay to four degenerate $\frac{1}{\sqrt{3}}(\pm 1,\pm 1,1)$ domains. After the domains have some time to coalesce and grow into larger regions, quantised fractional charges $Q=\pm\frac{1}{6}$ form wherever four different domains meet. There are always six more $\mathcal{Q}=-\frac{1}{6}$ than $\mathcal{Q}=\frac{1}{6}$ charges to preserve the initial $\mathcal{Q}=-1$. For example, panel (g) has nine $Q=-\frac{1}{6}$ and three $Q=\frac{1}{6}$ charges, while panel (d) fourteen $Q=-\frac{1}{6}$ and eight $Q=\frac{1}{6}$ charges. Note that we use periodic boundary conditions, so charges at the boundaries appear to be split into two pieces.
    }\label{fig:explodedSkyrmionNegKappa}%
\end{figure} 

In Fig.~\ref{fig:explodedSkyrmionNegKappa}, we show the results of a skyrmion explosion for the same quench and parameters as in Fig.~\ref{fig:decayingSkyrmion}, except $\kappa_c=-3.2$. A proliferation of $\mathcal{Q}=\pm\frac{1}{6}$ charges is observed, albeit with always six more $\mathcal{Q}=-\frac{1}{6}$ than $\mathcal{Q}=\frac{1}{6}$ charges to conserve the initial topological charge $\mathcal{Q}=-1$. As described in the figure caption, the system is unstable and more and more defects are nucleated.

\section{Ferromagnetic phase for the model in Eq.~(8)}\label{app:DomainWalls}
Here, we give the formula for the magnetisation in the ferromagnetic phase for the model defined in Eq. (8).
Substituting the parametrisation $\magvecunit=[\sin(\theta_{\text{FM}})\cos(\phi_{\text{FM}}),\sin(\theta_{\text{FM}})\sin(\phi_{\text{FM}}),\cos(\theta_{\text{FM}})]^T$ into Eq.~\eqref{eq:modelEasyPlaneChiralMagnet} and minimising with respect to $\theta_{\text{FM}}$, we obtain the solution shown in Eq. (9) of the main text with
\begin{align}\label{eq:thetaDefectFerromagnet}
\theta_{\text{FM}}(b,\kappa_p,\kappa_u)&=\arccos\left(\frac{-g^{\frac{2}{3}}-2\sqrt[3]{6}\kappa_p (\kappa_p-\kappa_u)}{6^{\frac{2}{3}} \kappa_p g^{\frac{1}{3}}}\right),\nonumber\\
g&=\sqrt{3} \sqrt{\kappa_p^3 \left(27 b^2 \kappa_p-16 (\kappa_p-\kappa_u)^3\right)}-9b\kappa_p^2.
\end{align}
As $b$ is varied between $0$ and $b_{\text{crit}}$, $\theta_{\text{FM}}(b,\kappa_p,\kappa_u)$ interpolates smoothly between the values $\frac{\pi}{2}$ and $\frac{\pi}{2}(1-\sign(b)$.

\section{Transport cross sections}\label{sec:scattering}


To compute quantities such as spin and thermal conductivities or forces on magnetic textures arising from magnon scattering, one has to keep track of the change in momentum of magnons when an incoming magnon with $\vb k_\text{in}=k (1,0)$ is scattered to  $\vb k_\text{out}=k (\cos \chi,\sin \chi)$. The rate of change of momentum is then given by
\begin{align}\label{eq:momentumTransfer}
  \diff{\Delta \vb  P}{t}=-j_m \, \hbar k \,\begin{pmatrix}\sigma_\| \\\sigma_\perp \end{pmatrix},
\end{align} where $j_m$ is the magnon current. The transport cross sections $\sigma_{\parallel}$ and $\sigma_\perp$ are computed from the differential cross section using
\begin{equation}
\begin{aligned}
\sigma_{\parallel}&={ \int}\dd{\chi}(1-\cos\chi)\frac{\mathrm{d}\sigma}{\mathrm{d}\chi}\thinspace,\\
\sigma_\perp&={ \int}\dd{\chi}(-\sin\chi)\frac{\mathrm{d}\sigma}{\mathrm{d}\chi}. \label{eq:TrCS}
\end{aligned}
\end{equation}
Eq.~\eqref{eq:momentumTransfer} is valid in a limit where magnon damping can be neglected, see App.~\ref{app:analytics} for a calculation in the presence of a finite Gilbert damping $\alpha$.

In the limit of large momentum, one can simply calculate the cross section from classical physics \cite{schroeter2015}.
Incoming charged particles with a high velocity are deflected very little, thus their trajectory is well approximated by $\vb r(t)=(v t,y_0)$. The total momentum transfer per particle is then given by
$\Delta \vb P=\int \dd{t} q (\vb v \times \vb B)$. Multiplying this with the current of incoming particles,  one obtains the following rate of change of momentum for $\vb B=(0,0,B(\vb{r}))$,
\begin{align}
\diff{\Delta P_\perp}{t}=- j_m \int \dd{y_0} \int \dd{t}\, q \, v \, B(\vb r(t))=-j_m \int \dd^2{r} \,q \,B(\vb r),
\end{align}
and therefore, in this large-$k$ limit, 
\begin{align}
\sigma_\perp \approx \frac{q \Phi}{\hbar k}=\frac{2 \pi (\Phi/\Phi_0)}{ k} \label{sigmaPerpClass},
\end{align}
where $\Phi$ is the total flux and $\Phi_0=2 \pi \hbar/q$ is the flux quantum. In contrast, $\sigma_\|$ is quadratic in the external field and vanishes within the approximation used above.

In the opposite limit of $k \to 0$, the wavelength is much larger than the size of the defect. Thus, one has to study the Aharonov-Bohm setup, where $B(\vb r)=\Phi \delta^2(\vb r)$. In this case, one obtains
\begin{align}
  \sigma_{\perp}&\approx\frac{\sin(2 \pi \Phi/\Phi_0)}{k} \label{eq:sigmaPerp0}.
\end{align}
The derivation of Eq. \eqref{eq:sigmaPerp0} is remarkably subtle and has been the source of major controversies in the literature 
\cite{wexler1998,nielsen1995,berry1999,keating2001,shelankov1998}. Technically, the problem arises from the treatment of the (highly singular) forward scattering. This problem can, e.g., be resolved by studying 
the scattering of finite-size wave packets or of a beam of incoming particles of finite width \cite{shelankov1998}. Alternatively, one can analyse the scattering problem of an incoming plane wave at {\em finite} distance from the scattering centre \cite{nielsen1995}.
In all these cases, it is essential to correctly include the interference of incoming and scattered waves for $\chi = 0$ to obtain the correct formula for $\sigma_\perp$ (there is no such problem from $\sigma_\|$). Only when the AB scattering problem is correctly regularised for $\chi \to 0$, does one find that the classical result, Eq.~\eqref{sigmaPerpClass},  and the quantum result, Eq.~\eqref{eq:sigmaPerp0},  coincide for $\Phi \to 0$.

The discussion given above assumed a charged particle coupled to a magnetic field $(0,0,B(\vec r))$. We have, however, also considered other contributions. Magnon number is not a conserved quantity; therefore, magnons can be absorbed, or a high-energy magnon can decay into a pair of low-energy magnons. In the limit of vanishing Gilbert damping, $\alpha \to 0$, magnon absorption is forbidden by energy conservation. Inelastic decay is also not possible by energy conservation if the magnon energy $E_m$ is in the energy window $\hbar\Omega_g < E_m < 2 \hbar\Omega_g$. We can thus ignore magnon absorption in the small-$\alpha$ limit, when the magnon mean-free path is large compared to the size of the domain, see App~\ref{app:analytics} for an analytical calculation of magnon currents for finite $\alpha$.

Besides the effects from $B(\vec r)$, one has to take into account that the presence of the magnetic texture will lead to an effective potential, $V(\vec r)$, for magnons \cite{SchuetteScatt2014}. The classical result Eq.~\eqref{sigmaPerpClass} is not affected by  $V(\vec r)$ in the large-$k$ (classical) limit, as in this case the kinetic energy is large compared to $V(\vec r)$. The small-$k$ limit, Eq.~\eqref{eq:sigmaPerp0}, is more subtle. Here, the cross section arises from the interference of magnons passing the defect far away from the defect centre. However, far from the defect centre the magnons have to cross a domain wall, see Fig. 4(a).
Only magnons which are transmitted through the domain wall can contribute to the Aharonov-Bohm effect and thus to the transverse force on the defect. Therefore, we multiply the formula for the cross section in Eq.~\eqref{eq:sigmaLimits} by the transmission rate $T_k$ through the domain wall, which may vanish for $k\to 0$.

\section{Fractional flux driven by magnon currents: numerics}\label{app:NumericsDefect}

For our numerical micromagnetic simulations, we use the GPU-accelerated open-source software mumax3 \cite{vansteenkiste2014} in combination with mathematica \cite{mathematica2023}.  

\begin{figure}
\includegraphics[width=0.3\textwidth]{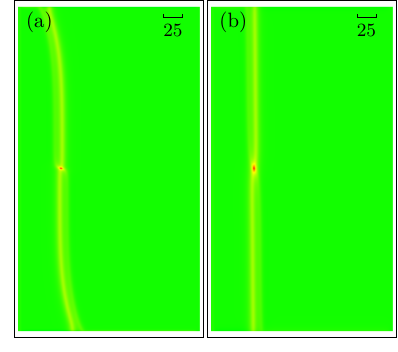}%
	\caption{Fractional defect with (a) $A_{\text{ani}}=0$ and (b) $A_{\text{ani}}=10$, other parameters as in Fig.~\ref{fig:easyplaneDMI}. Note that here we show the full system with $200\times1000$ spins, which results in a different aspect ratio. For greater clarity, we only show the $\hat{M}_z$ data. The system with $A_{\text{ani}}=0$ has curved domain wall edges at the top and bottom, which drag the defect sideways when a magnon current is applied from the left.
	}\label{fig:FractionalDefectJani}%
\end{figure}

\textit{Generation of the defect ---} To generate the defect in Fig.~\ref{fig:easyplaneDMI}, we start with a grid of width $N_x=100$ and height $N_y=1000$ cells, with lattice spacing $\Delta=1.25$. We fix the spins along the right and left edges to take the values $\left(\pm\sin(\theta_{\text{FM}})/\sqrt{2},\pm\sin(\theta_{\text{FM}})/\sqrt{2},\cos(\theta_{\text{FM}})\right)^T$, respectively, calculating $\theta_{\text{FM}}$ using Eq.~\ref{eq:thetaDefectFerromagnet} for the parameters $\kappa_p=0.05,\kappa_u,b=-0.006$. We turn off dipole-dipole interactions and set the Heisenberg coupling $A=1$ and the interfacial Dzyaloshinskii-Moriya interaction (DMI) $D=0.09$. We also add a custom-made (see supplementary go file \texttt{drivenDefect.go} lines 116--127 in \cite{zenodo2023} for the technical implementation) anistropic Heisenberg coupling $A_{\text{ani}}=10$. The purpose of $A_{\text{ani}}$ is to stiffen the system in the $y$-direction, as the domain wall otherwise has a tendency to deform (becoming diagonal) once we start driving it with a magnon current, see Fig.~\ref{fig:FractionalDefectJani}(a). We found that the addition of $A_{\text{ani}}$, as well as keeping $b$ close to zero, keeps the domain walls nicely vertical throughout the entire duration of the driven simulation. By letting the system relax, we generate the ``domain wall within a domain wall'', at the centre of which a topological defect of $\mathcal{Q}=-0.352$ forms. We then import the magnetisation .ovf file into mathematica and add one hundred copies of the rightmost column of spins to the right, thereby enlarging the system to be $N_x=200$ wide and $N_y=1000$ long. The purpose of this is to minimise magnon reflection from the right edge of the system back onto the defect, which can otherwise cause unwanted interference effects. Note that we calibrated this length for our value of Gilbert damping $\alpha=0.03$ --- for smaller values of $\alpha$, more copies would need to be placed to the right of the defect to ensure that the magnons have sufficiently decayed and won't be reflected back. Exporting the enlarged .ovf file back to mumax3, we relax the system once more before proceeding to the next stage of driving it with a magnon current.


\textit{Driving with a magnon current \& measuring the defect velocity ---} We start injecting magnons with frequency $\omega$ and wavevector $\vb{k}=k\vb{e}_x$ into the system by forcing the magnetisation of all the spins along the left edge of the sample to oscillate with amplitude $\delta\vb{M}(t)=2\delta \magvec \cos(\Omega t)$, where $\delta\vb{M}\perp \vb{M}^{(0)}$, and $\vb{M}^{(0)}$ is the equilibrium magnetisation. To track the position of the defect in time, we define the defect coordinate $\vb{R}$ as the centre of mass of the topological charge distribution, Fig.~\ref{fig:easyplaneDMI}(b),
\begin{equation}\label{eq:centreOfmassNumerics}
\vb{R}=\frac{1}{\mathcal{Q}}\int\dd^2{r}\,\vb{r}\,\rho_{\text{top}}(\vb{r}).
\end{equation}
To calculate Eq.~\eqref{eq:centreOfmassNumerics} using our discrete numerical data, we change $\int\dd^2{r}\to \Delta^2\sum_{i,j},\vb{r}\to \Delta(i\vb{e}_x+j\vb{e}_y)$ and approximate the derivatives in $\rho_{\text{top}}$ by $\nabla_i\magvecunit=\frac{1}{\Delta}\left(\magvecunit(\vb{r}+\Delta\hat{\vb{e}}_i)-\magvecunit(\vb{r})\right)$. After waiting a few hundred oscillations for the steady state to be established, we measure the defect velocity $\vb{V}=\dot{\vb{R}}$. We repeat this procedure for different driving frequencies $\Omega$ to investigate the low- and high-$k$ scattering limits, obtaining the results plotted in Fig.~\ref{fig:easyplaneDMI}(c).


\section{Fractional flux driven by magnon currents: analytics}\label{app:analytics}
In this section, we analytically derive, for a strongly simplified toy model how the force on a magnetic texture can be obtained via perturbation theory directly from the LLG equation. As described below, we study a simplified rotationally invariant magnon scattering problem, where an effective magnetic flux is added ``by hand'', see below for details.
The analytical calculation can be used to check some of the  assumptions underlying  Eq.~\eqref{eq:forceTransverseMagnonCurrent}, including the effects of a 
small Gilbert damping.


For the sake of our calculation, it is mathematically more convenient to consider the dynamics in our system as being induced by an oscillating external magnetic field $\vb{B}_1(\vb{r},t)$, where $B_1\sim\delta M/M_0$. In the limit of weak driving, $\epsilon=\delta M/M_0\ll 1$, the following perturbative ansatz accurately describes \cite{delser2023} the time-dependent magnetisation,
\begin{equation}\label{eq:magvecPertExpansion}
\magvecunit(\vb{r},t)=\magvecunit^{(0)}(\vb{r}-\epsilon^2\vb{V}\,t)+\epsilon\vb{M}^{(1)}(\vb{r})+\mathcal{O}(\epsilon^2),
\end{equation}
where $\magvecunit^{(0)}$ is the equilibrium (normalised) magnetisation, $\magvec^{(1)}$ is the linear response and $\vb{V}$ is the constant linear velocity of the magnetic texture. As has been discussed in Ref.~\cite{delser2023}, it is sufficient to compute oscillations of the magnetisation linearly in $\epsilon$ to determine the velocity to order $\epsilon^2$. Substituting Eq.~\eqref{eq:magvecPertExpansion} into the Landau-Lifshitz-Gilbert equation, crossing with $\magvec$, projecting onto $\nabla\magvec$ and integrating over space, see \cite{delser2023} App.B for technical details, we arrive at an effective Thiele \cite{thiele1973} equation,
\begin{equation}\label{eq:thieleEq}
\vb{G}\cross\vb{V}+\alpha \mathcal{D}\vb{V}=\vb{F},
\end{equation}
where $\vb{G}=\vb{e}_z\frac{M_0}{|\gamma|}4\pi \mathcal{Q}$ is the gyrocoupling vector, $\mathcal{D}_{ij}=\frac{M_0}{|\gamma|}\int\dd^2{r}\,\nabla_i\magvecunit\cdot\nabla_i\magvecunit$ is the dissipation matrix, $\gamma=qg/(2m)$ is the gyromagnetic ratio and $\vb{F}$ is the force divided by the sample thickness, defined by
\begin{equation}\label{eq:forceYjellyfish}
\begin{aligned}
F_i=&\left(\frac{\delta M}{M_0}\right)^2\frac{M_0}{|\gamma|}\Big\langle \int\dd^2{r}[\magvecunit^{(0)}\cdot(\dot{\magvec}^{(1)}\cross\nabla_i\magvec^{(1)})\\
&+\alpha \dot{\magvec}^{(1)}\cdot \nabla_i\magvec^{(1)}-|\gamma|\vb{B}_1\cdot\nabla_i\magvec^{(1)}]\Big\rangle_{T}.
\end{aligned}
\end{equation}
The time averaging over one period of the drive, $T=2\pi/\Omega$, gets rid of any oscillating terms and ensures that $F_i$ is constant in time.

Let us consider Eq.~\eqref{eq:thieleEq} specifically for the ``domain wall within a domain wall'' system, Fig.~\ref{fig:easyplaneDMI}(a). While $\mathcal{D}_{xx}\sim L_y$ scales proportionally to the  length of the system in the $y$-direction, the other components of the dissipation matrix --- $\mathcal{D}_{xy},\mathcal{D}_{yy}$, and $G$ --- are independent of $L_y$. Similarly, there is one linear in $L_y$ contribution to the force $F_x$, which arises from the reflection of magnons from the domain wall far away from the point defect. While this nominally leads to a velocity $V_x \approx F_x/(\alpha D_{xx})$, we find numerically that this effect is very small. Therefore, we neglect it in the following discussion. All other forces are independent of $L_y$. Thus,
Eq.~\eqref{eq:thieleEq} simplifies to
\begin{equation}\label{eq:velocitiesDefectTheory}
V_y\approx \frac{F_y}{\alpha \mathcal{D}_{yy}},
\end{equation}
i.e., the defect velocity is proportional to the transverse scattering force $F_y$. As $F_x$ does not affect $\vb{V}$ in this limit, we don't need to bother calculating it.



If we now wanted to calculate $F_y$ in full glory, we would need to solve for $\magvec^{(1)}$ around the equilibrium magnetisation $\magvecunit^{(0)}$ which describes the defect texture in Fig.~\ref{fig:easyplaneDMI}(a). The absence of any rotational symmetry (unlike, e.g., for a skyrmion or a vortex) makes this a very difficult problem to treat analytically. Instead, we will considerably simplify the task by solving a much easier problem, which simultaneously still retains the primary physical feature we are interested in, namely, the influence of the fractional magnetic flux, but is also rotationally symmetric.

The simplified problem statement is the following: We assume that we have an infinite ferromagnetic 2D system with the same free energy as the system with the defect, Eq.~\eqref{eq:modelEasyPlaneChiralMagnet}, but where instead $\magvecunit^{(0)}$ points parallel to the spins along the left boundary in Fig.~\ref{fig:easyplaneDMI} everywhere in the 2D plane. In order to switch to polar coordinates, we need to make the Heisenberg coupling in  Eq.~\eqref{eq:modelEasyPlaneChiralMagnet} isotropic, which is easily achieved by using the rescaled $y$-coordinate $y\to\sqrt{\frac{A}{A+A_{\text{ani}}}}y$. This rescaling has no impact on either the total topological charge $\mathcal{Q}$ or $F_y$, as it leaves the combination $\dd{y}\nabla_y$ carried by both these quantities unchanged. We place the flux resulting from the topological charge at the centre of our coordinate system, modelling it for example by $\Phi(r)=\Phi(1-e^{-r^2/R^2})\Phi_0$ (the precise shape is not important for the calculation, as long as it retains rotational symmetry). We parametrise 
\begin{equation}\label{eq:magnetisationParametrisation}
\begin{aligned}
\magvec^{(1)}(\vb{r},t)&=\vb{e}_-a(\vb{r},t)+\vb{e}_+a^*(\vb{r},t),\\
|\gamma|\vb{B}_1(t)&=b_1(\vb{r},t)\vb{e}_-+b_1^*(\vb{r},t)\vb{e}_+,
\end{aligned}
\end{equation}
where $a(\vb{r},t)$ are the linear response excitation fields and $\vb{e}_{\pm}$ are complex vectors perpendicular to $\magvecunit^{(0)}$, defined by 
\begin{equation*}
\vb{e}_{\mp}=\frac{1}{\sqrt{2}}\begin{pmatrix}
-\frac{1}{\sqrt{2}}\cos(\theta_{\text{FM}}) \mp \frac{i}{\sqrt{2}} \\
-\frac{1}{\sqrt{2}}\cos(\theta_{\text{FM}}) \pm \frac{i}{\sqrt{2}}  \\
-\sin(\theta_{\text{FM}})
\end{pmatrix}.
\end{equation*}
Using the Fourier conventions
\begin{equation}\label{eq:fourierConvention}
\begin{aligned}
a(\vb{r},t)&=e^{i\Omega t}a_{+\Omega}(\vb{r})+e^{-i\Omega t}a_{-\Omega},(\vb{r}),\\
a_{\pm\Omega}(\vb{r})&=\sum_me^{im\chi}a_{\pm\Omega,m}(r),\\
b_1(t)&=b_{+\Omega}e^{i\Omega t}+b_{-\Omega}e^{-i\Omega t},
\end{aligned}
\end{equation}
the equation of motion for the Fourier-transformed fields $a_{\pm\omega}(\vb{r})$ is reduced to a time-independent single-variable differential matrix equation
\begin{equation}\label{eq:timeIndepEoMlinearResponse}
\begin{aligned}
&\sum_m e^{im\chi}\Omega(-1+i\alpha\sigma^z)\begin{pmatrix}a_{+\Omega,m}\\ a^*_{-\Omega,-m}\end{pmatrix}=\\
&-\sum_m e^{im\chi}\sigma^z\begin{pmatrix}D(r) & 2B\\ 2B & D(r)\end{pmatrix}\begin{pmatrix}a_{+\Omega,m}\\ a^*_{-\Omega,-m}\end{pmatrix}\\
&+\sigma^z\begin{pmatrix}b_{+\Omega}(\vb{r}) \\ b^*_{-\Omega}(\vb{r})\end{pmatrix},
\end{aligned}
\end{equation}
where
\begin{equation*}
\begin{aligned}
    D(r)&=\frac{|\gamma|}{M_0}\Bigg[-A\left(\frac{\partial^2}{\partial r^2}+\frac{1}{r}\frac{\partial}{\partial r}-\frac{1}{r^2}\left(m-\frac{\Phi(r)}{\Phi_0}\right)^2\right)\\
    &+\frac{1}{8} (8 b_0 \cos(\theta_{\text{FM}})-4 (\kappa_p+3 \kappa_u) \cos(2 \theta_{\text{FM}})\\
    &-5 \kappa_p \cos(4 \theta_{\text{FM}})+9 \kappa_p-4 \kappa_u)\Bigg],\\
    B&=\frac{|\gamma|}{M_0}\frac{1}{4} \sin^2(\theta_{\text{FM}})\left(3 \kappa_p \cos (2 \theta_{\text{FM}})-3 \kappa_p+2 \kappa_u\right)
\end{aligned}
\end{equation*}
are the coefficients of the $a^*_{\pm\Omega,m}a_{\pm\Omega,m}$ and Bogoliubov-like $a^*_{\pm\Omega,m}a_{\mp\Omega,-m}$ terms in $\mathcal{F}$, respectively. Note that we have suppressed the radial dependency of the $a_{\pm\Omega,m}$ fields to make the notation neater, although it is implied everywhere. For more details on the derivation of Eq.~\eqref{eq:timeIndepEoMlinearResponse}, see Sec.4 and App.F in \cite{delser2023}. 

To simplify Eq.~\eqref{eq:timeIndepEoMlinearResponse}, we use the following ansatz
\begin{equation}\label{eq:ansatzLinearResponseMagnonFlux}
    \begin{pmatrix} a_{+\Omega,m}(r) \\ a^*_{-\Omega,-m}(r)\end{pmatrix}=\int_{-\infty}^{\infty} \frac{\mathop{dk}}{2\pi}\begin{pmatrix} u_k \\v_k\end{pmatrix}c_{m,k}\psi_m(kr),
\end{equation}
where $\psi_m(kr)$ is the solution of the differential equation
\begin{equation*}
    -\left(\frac{\partial^2}{\partial r^2}+\frac{1}{r}\frac{\partial}{\partial r}-\frac{1}{r^2}\left(m-\frac{\Phi(r)}{\Phi_0}\right)^2\right)\psi_m(kr)=k^2\psi_m(kr).
\end{equation*}
Substituting Eq.~\eqref{eq:ansatzLinearResponseMagnonFlux} into Eq.~\eqref{eq:timeIndepEoMlinearResponse}, we see that the $r$-dependent part of $D(r)$ simplifies to an $r$-independent prefactor, consequently we can relabel the resulting combination $D(r)\to D(k)=\frac{|\gamma|}{M_0}Ak^2$. We pick $(u_k,v_k)^T$ to satisfy the following eigenvalue equation,
\begin{equation*}
\Omega_k(-1+i\alpha\sigma^z)\begin{pmatrix}
    u_k \\ v_k
\end{pmatrix}=-\sigma^z\begin{pmatrix}
    D(k) & 2B \\ 2B & D(k)
\end{pmatrix}\begin{pmatrix}
    u_k \\ v_k
\end{pmatrix},
\end{equation*}
where the eigenfrequency $\Omega_k$ is complex owing to the finite damping $\alpha$. Here, we only keep the $\Omega_k$ whose real part is positive, as we keep the driving frequency $\Omega>0$, although there is formally also a negative eigenfrequency with $\text{Re}[\Omega_k]<0$. Eq.~\eqref{eq:timeIndepEoMlinearResponse} then simplifies to
\begin{equation}\label{eq:EoMlinearAnsatz}
\begin{aligned}
&\sum_{m}e^{im\chi}(-1+i\alpha\sigma^z)\int_0^{\infty} \frac{k\mathop{dk}}{2\pi}(\Omega-\Omega_k)\begin{pmatrix} u_k \\v_k\end{pmatrix}c_{m,k}\psi_m(kr)\\
&=\sigma^z\begin{pmatrix}b_{+\Omega}(\vb{r}) \\ b^*_{-\Omega}(\vb{r}) \end{pmatrix}.
\end{aligned}
\end{equation}
 Eq.\ref{eq:EoMlinearAnsatz} lets us calculate either $c_{m,k}$ if $b_{\pm\Omega}(\vb{r})$ is known, or vice versa. In our case, we choose the $c_{m,k}$ to be
 \begin{equation}\label{eq:cCoeff}
 c_{m,k}=\frac{1}{i}\frac{|v_g(k)|\lambda(k)}{\Omega-\Omega_k}i^{|m|}e^{i\delta_{m,k}}e^{ikL_x},
 \end{equation}
where $v_g(k)=\frac{d\Omega_k}{dk}$ and $\lambda(k)$ is an even function without any simple poles defined in such a way that the boundary condition 
\begin{equation}\label{eq:BoundaryCond}
a_{\pm\Omega}(x=-L_x)=1,
\end{equation}
defined along the left edge of the system in our numerics, is satisfied. The denominator $\Omega-\Omega_k$ has four simple poles $\pm k_t,\pm k_{\text{eva}}$ in the complex $k$ plane, where $k_t,k_{\text{eva}}$ represent the travelling and evanescent solutions, respectively.  $k_t$ takes the form $k_{t}=k_0-i\alpha\Omega/v_g$, where the real part $k_0\sim \sqrt{\Omega-\Omega_g}$ is the momentum of the magnon, while the imaginary part $\alpha\Omega/v_g$, with $v_g=d\Omega/dk$, describes the damping. $k_{\text{eva}}$ is mostly negative imaginary at all driving frequencies above the gap. To verify that our choice of $c_{m,k}$ correctly reproduces the scattering ansatz of incident plane wave plus scattered radial wave in the far-field region, we substitute Eq.~\eqref{eq:cCoeff} into Eq.~\eqref{eq:ansatzLinearResponseMagnonFlux}, followed by Eq.~\eqref{eq:fourierConvention}, and use the asymptotic form of $\psi_m(kr)$ in the far field, $\lim_{r\gg R}\psi_m(kr)=\sqrt{\frac{2}{\pi k r}}\cos\left(kr-\frac{\pi|m|}{2}-\frac{\pi}{4}+\delta_m\right)$, to rewrite it as
\begin{align}
   & \lim_{r\gg R}\begin{pmatrix} a_{+\Omega}(\vb{r}) \\ a^*_{-\Omega}(\vb{r})\end{pmatrix}=\nonumber\\
   &\int_{-\infty}^{\infty} \frac{\mathop{dk}}{2\pi i}\begin{pmatrix} u_k \\v_k\end{pmatrix}\frac{|v_g(k)|\lambda(k)}{\Omega-\Omega_k}e^{ikL_x}\cdot \\
   &\sum_me^{im\chi}\left(i^{m}J_{m}(kr)+\frac{1}{2}(e^{2i\delta_{m,k}}-1)H^{(1)}_{0}(kr)\right),\nonumber
\end{align}
where $H^{(1)}_{n}(kr)\sim \sqrt{\frac{2}{\pi k r}}e^{i(kr-\frac{n\pi}{2}-\frac{\pi}{4})}$ (also valid for complex $k$) is the Hankel function of the first kind. Using the Bessel expansion of a plane wave, we can replace $\sum_m i^mJ_m(kr)e^{im\chi}=e^{ikx}$. Then, transforming the integral over $k$ into an integral over the complex $k$ plane, where we close the semi-circle in the upper/lower contours depending on the signs of $L_x+x,L_x\pm r$, we arrive at the following expression
\begin{equation}\label{eq:scattAnsatz}
\begin{aligned}
   & \lim_{r\gg R}\begin{pmatrix} a_{+\Omega}(\vb{r}) \\ a^*_{-\Omega}(\vb{r})\end{pmatrix}=\\
   &\lambda(k_t)e^{-ik_t|L_x+x|}\begin{pmatrix} u_{k_t} \\v_{k_t}\end{pmatrix}+\lambda(k_{\text{eva}})e^{-ik_{\text{eva}}|L_x+x|}\begin{pmatrix} u_{k_{\text{eva}}} \\v_{k_{\text{eva}}}\end{pmatrix}\\
   &-ie^{-\frac{i\pi}{4}}f_{k_0}(\chi)\lambda(k_t)\frac{e^{-ik_t(L_x+r)}}{\sqrt{r}}\begin{pmatrix} u_{k_t} \\v_{k_t}\end{pmatrix} + \mathcal{O}(\alpha),
\end{aligned}
\end{equation}
where the net scattering amplitude $f_k(\chi)=\frac{1}{\sqrt{2\pi k}}\sum_m(e^{2i\delta_{m,k}}-1)e^{im\chi}$. Note that in the derivation of Eq.~\eqref{eq:scattAnsatz}, we used $\delta_{m,-k}=\delta_{m,k}$, and also did not keep the $e^{-ik_{\text{eva}}(L_x+r)}$ term as it is vanishingly small. To match the boundary condition Eq.~\eqref{eq:BoundaryCond}, we must have
\begin{equation}\label{eq:BoundCondEq}
\lambda_{k_t}\begin{pmatrix} u_{k_t} \\v_{k_t}\end{pmatrix}+\lambda_{k_{\text{eva}}}\begin{pmatrix} u_{k_{\text{eva}}} \\v_{k_{\text{eva}}}\end{pmatrix}=\begin{pmatrix} 1 \\ 1 \end{pmatrix},
\end{equation}
where we neglected the scattered radial wave term. This is a justified approximation as long as the damping $\alpha$ and distance $L_x$ to the defect are large enough that the scattered wave has significantly decayed by the time it reaches $x=-L_x$, with $\exp(-2\alpha \Omega L_x/v_g)\ll 1$. We can then solve Eq.~\eqref{eq:BoundCondEq} for $\lambda_{k_t}$, $\lambda_{k_{\text{eva}}}$ as a function of $\Omega$. For our set of parameters $b,\kappa_p,\kappa_u$, this results in $\lambda_{k_0},\lambda_{k_{\text{eva}}}$ interpolating between $\lambda_{k_0}=1.12$ and $\lambda_{k_{\text{eva}}}=0$ at $\Omega=\Omega_g$ and $\lambda_{k_0}=\lambda_{k_{\text{eva}}}=1$ at high values $\Omega\gg \Omega_g$.

Eq.~\eqref{eq:scattAnsatz} also lets us evaluate $\vb{B}^{(1)}(\vb{r},t)$ in the far field. By substituting Eq.~\eqref{eq:scattAnsatz} into Eq.~\eqref{eq:timeIndepEoMlinearResponse}, we see that $\vb{B}^{(1)}(\vb{r},t)$ is zero everywhere except where $\nabla_i\magvec^{(1)}$ has discontinuities, i.e. along the $x=L_x$ line. Integrating Eq.~\eqref{eq:timeIndepEoMlinearResponse} between $x=-L_x\mp\epsilon$, we obtain
\begin{equation}
\begin{aligned}
&\lim_{r\gg R}
\begin{pmatrix}
b_{+\Omega}(\vb{r}) \\ b^*_{-\Omega}(\vb{r}) \end{pmatrix}=\delta(x+L_x)\cdot \\
&\frac{2iA|\gamma|}{M_0}\left(
k_t\lambda(k_t)
\begin{pmatrix}
u_{k_{t}} \\ v_{k_{t}}
\end{pmatrix} 
+ k_{\text{eva}}\lambda(k_{\text{eva}})
\begin{pmatrix}
u_{k_{\text{eva}}} \\ 
v_{k_{\text{eva}}}\end{pmatrix} 
\right).
\end{aligned}
\end{equation}

Having fully solved for $\magvec^{(1)}(\vb{r},t)$ and $\vb{B}^{(1)}(\vb{r},t)$ in our simplified toy model, we can finally substitute it into Eq.~\eqref{eq:forceYjellyfish} and calculate $F_y$. The first term in the integrand can be written as a full derivative and therefore gives zero contribution to $F_y$ after integration (see also App. J in \cite{delser2023} for more details). Physically, this arises because this term describes the magnon current, which vanishes at $r=\infty$. The third term with $\vb{B}_1(t)$ also vanishes, as $\vb{M}^{(1)}(-L_x,y\to\infty)=\vb{M}^{(1)}(-L_x,y\to -\infty)$, from Eq.~\eqref{eq:scattAnsatz}. Therefore, any non-zero contribution can only come from the second term, proportional to $\alpha\langle\int\magvec^{(1)}\cdot\nabla_y\magvec^{(1)}\rangle_T$. As we are interested in $F_y$ in the limit of small damping $\alpha\to 0$, we only keep those terms coming from the integral which scale like $1/\alpha$, as these cancel the $\alpha$ prefactor to give an overall $\alpha$-independent contribution to $F_y$. The only way to get such terms is for the $\nabla_y$ to act on the exponential factor $e^{-ik_t(L_x+r)}$, resulting in an overall term $e^{-\frac{2\alpha v_g}{\Omega}(L_x+r)}$ in the integrand, which scales like $1/\alpha$ after integration over $r$. Thus, we obtain
\begin{equation}
 \begin{aligned}
F_y&=\frac{\delta M^22\alpha\Omega}{M_0|\gamma|} \, \text{Im}\, \int \dd^2{r}\Bigg(a^*_{+\Omega}(\vb{r})\nabla_y a_{+\Omega}(\vb{r})\\
&\quad\quad-a^*_{-\Omega}(\vb{r})\nabla_y a_{-\Omega}(\vb{r})\Bigg)\\
&\approx \frac{\delta M^22\alpha\Omega}{M_0 |\gamma|} e^{-\frac{2\alpha\Omega L_x}{v_g}}|\lambda(k_t)|^2(|u_k|^2+|v_k|^2)\cdot\\
&\quad\int_R^{\infty} r\dd{r} \frac{e^{-\frac{2\alpha\Omega r}{v_g}}}{r}\int_0^{2\pi} \dd{\chi}(-\sin\chi)\frac{\dd{\sigma}}{\dd{\chi}}\\
&\approx\frac{|\lambda(k_t)|^2\delta M^2}{M_0|\gamma|} v_g e^{-\frac{2\alpha\Omega}{v_g}L_x}k\sigma_{\perp},
    \end{aligned}
\end{equation}
where in the fourth line we used the definition of the differential cross section, $\frac{d\sigma}{d\chi}=|f_k(\chi)|^2$. Comparing this with Eq.~\eqref{eq:forceTransverseMagnonCurrent}, we identify
\begin{equation}\label{eq:magnonCurrentPrecise}
    j_m\approx\frac{M_0}{\hbar|\gamma|}\left(\frac{\delta M}{M_0}\right)^2v_g(\Omega)|\lambda(k_t)|^2e^{-\frac{2\alpha\Omega}{v_g}L_x}.
\end{equation}
For finite $\alpha$, $j_m$ drops to zero when $\Omega\to \Omega_g$, as $v_g\to 0$, which explains why $F_y$, and therefore also $V_y$, are zero close to the gap, see Fig.~\ref{fig:easyplaneDMI}(c) (note that the notation in the figure is different as we are using rescaled $v_y=V_y/|\gamma|$, $\omega=\Omega/|\gamma|$ there). The same exponential factor implies that $j_m$ should vanish in the large-$\Omega$ limit. By construction, Eq.~\eqref{eq:magnonCurrentPrecise} also tells us that $j_m$ and therefore $F_y$ depend quadratically on $\delta M/M_0$, which we also checked numerically (data not shown).

\end{document}